\documentclass[pdflatex,sn-nature]{sn-jnl}

\usepackage{graphicx}%
\usepackage{multirow}%
\usepackage{amsmath,amssymb,amsfonts}%
\usepackage{amsthm}%
\usepackage{mathrsfs}%
\usepackage[title]{appendix}%
\usepackage{xcolor}%
\usepackage{textcomp}%
\usepackage{manyfoot}%
\usepackage{booktabs}%
\usepackage{algorithm}%
\usepackage{algorithmicx}%
\usepackage{algpseudocode}%
\usepackage{listings}%

\usepackage{upgreek}
\usepackage{lmodern}
\usepackage{anyfontsize}
\usepackage{gensymb}
\def\micron{$\upmu$m}
\def\mj{\ensuremath{\textrm{M}_J}}

\usepackage[sf]{titlesec}
\begin{document}

\title{A temperate super-Jupiter imaged with JWST in the mid-infrared}

\author*[1]{E.~C.~Matthews}\email{matthews@mpia.de}

\author[2]{A.~L.~Carter}

\author[3]{P.~Pathak}

\author[4]{C.~V.~Morley}

\author[5]{M.~W.~Phillips}

\author[6]{S.~Krishanth P.~M.} 

\author[7]{F.~Feng}


\author[8,9]{M.~J.~Bonse}

\author[1]{L.~A.~Boogaard}

\author[10]{J.~A.~Burt}

\author[11]{I.~J.~M.~Crossfield}

\author[6]{E.~S.~Douglas}

\author[1]{Th.~Henning}

\author[6]{J.~Hom}

\author[6]{C.-L.~Ko}

\author[12]{M.~Kasper}

\author[13,14]{A.-M.~Lagrange}

\author[15]{D.~Petit dit de la Roche}

\author[13]{F.~Philipot}

\affil[1]{\small Max Planck Institute for Astronomy, K{\"o}nigstuhl 17, D-69117 Heidelberg, Germany}
\affil[2]{\small Space Telescope Science Institute, 3700 San Martin Drive, Baltimore, MD 21218, USA}
\affil[3]{\small Department of SPASE, Indian Institutes of Technology Kanpur, 208016, UP, India}
\affil[4]{\small Department of Astronomy, University of Texas at Austin, 2515 Speedway, Austin, TX 78712, USA}
\affil[5]{\small Institute for Astronomy, University of Edinburgh, Royal Observatory, Blackford Hill, Edinburgh EH9 3HJ, UK}
\affil[6]{\small Steward Observatory and Department of Astronomy, 933 N Cherry Ave, Tucson, AZ, USA, 85721}
\affil[7]{\small Shanghai Jiao Tong University, Shengrong Road 520, Shanghai, 201210, People’s Republic Of
China}
\affil[8]{\small ETH Zurich, Institute for Particle Physics \& Astrophysics, Wolfgang-Pauli-Str. 27, 8093 Zurich, Switzerland}
\affil[9]{\small Max Planck Institute for Intelligent Systems, Max-Planck-Ring 4, D-72076 Tübingen, Germany}
\affil[10]{\small Jet Propulsion Laboratory, California Institute of Technology, 4800 Oak Grove Drive, Pasadena, CA 91109, USA}
\affil[11]{\small Department of Physics and Astronomy, University of Kansas, Lawrence, KS, USA}
\affil[12]{\small European Southern Observatory, Garching bei München, Germany}
\affil[13]{\small LESIA, Observatoire de Paris, Université PSL, CNRS, 5 Place Jules Janssen, 92190 Meudon, France}
\affil[14]{\small Univ. Grenoble Alpes, CNRS-INSU, Institut de Planétologie et d’Astrophysique de Grenoble (IPAG), UMR 5274, Grenoble 38041,
France}
\affil[15]{\small Department of Astronomy, University of Geneva, Chemin Pegasi 51, CH-1290 Versoix, Switzerland}

\maketitle

\section*{Abstract}

\textbf{Of the $\sim$25 directly imaged planets to date, all are younger than 500Myr and all but 6 are younger than 100Myr\citep{NasaExoArc}. Eps Ind A (HD209100, HIP108870) is a K5V star of roughly solar age (recently derived as 3.7-5.7Gyr\citep{Feng2019} and 3.5$^{+0.8}_{-1.3}$Gyr\citep{Chen2022}). A long-term radial velocity trend\citep{Endl2002,Zechmeister2013} as well as an astrometric acceleration\citep{Brandt2018,Kervella2019} led to claims of a giant planet\citep{Feng2019,Feng2023,Philipot2023} orbiting the nearby star (3.6384±0.0013pc\citep{gaia_dr3}). Here we report JWST coronagraphic images that reveal a giant exoplanet which is consistent with these radial and astrometric measurements, but inconsistent with the previously claimed planet properties. The new planet has temperature $\sim$275K, and is remarkably bright at 10.65\micron\ and 15.50\micron. Non-detections between 3.5-5\micron\ indicate an unknown opacity source in the atmosphere, possibly suggesting a high metallicity, high carbon-to-oxygen ratio planet. The best-fit temperature of the planet is consistent with theoretical thermal evolution models, which are previously untested at this temperature range. The data indicates that this is likely the only giant planet in the system and we therefore refer to it as ``b", despite it having significantly different orbital properties than the previously claimed planet ``b".}

\section*{Main}
\label{main}

We observed Eps~Ind~A with the JWST Mid-InfraRed Instrument (MIRI)\citep{Rieke2015} coronagraph on 2023-07-03, using two narrowband filters (10.65\micron\ and 15.50\micron). Figure 1 shows a bright point source detected in the north-east quadrant, at a separation of 4.11'': this is the opposite quadrant than expected based on previous orbital solutions\citep{Feng2019,Feng2023,Philipot2023}. The source is unresolved, has apparent magnitude 13.16 and 11.20 at 10.65\micron\ and 15.50\micron, and is consistent with a cold, Jupiter-sized object at the host star distance. We rule out chance-aligned background objects and recover the companion at low signal-to-noise in archival observations, confirming that it is physically associated with the host star. The source has significantly different orbital properties than the literature solutions for Eps~Ind~Ab, but we conclude that it is likely the only massive planet in the system and is responsible for the previously observed radial velocity and astrometric acceleration of the host star.

\begin{figure}
    \centering
    \includegraphics[width=\textwidth]{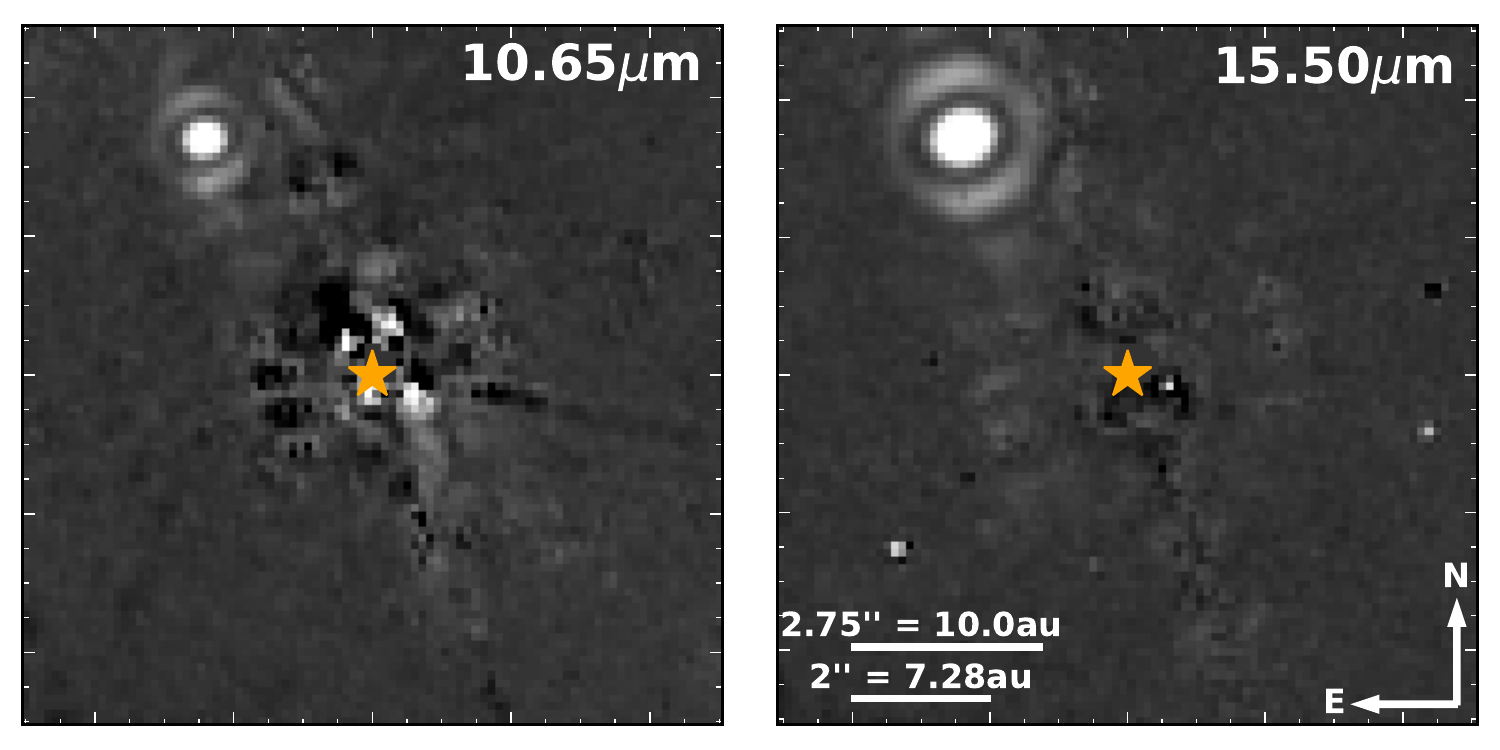} 
    \caption{\textbf{A point source is detected in JWST/MIRI coronagraphic images of Eps~Ind~A.} The target is observed at 10.65\micron\ and 15.50\micron\, and starlight is removed with reference differential imaging. Only the central portion of the field of view is shown, and the stellar position is marked with an orange star. A bright point source is detected in the upper left corner of these images, at a projected separation 4.11'' (15.0au at the distance of Eps~Ind~A)}
    \label{fig:jwstdata}
\end{figure}

Due to the unexpected location of the source, we first assess whether it could be a chance-aligned background object. This is statistically unfavored: source-counting studies with the MIRI broadband filters\citep{Wu2023} indicate an expected background density of $\sim$45 sources per square degree at least as bright as the point source -- corresponding to just a 0.027\% likelihood of finding such a source within 5'' of the target. While occurrence rates for giant, long-period planets are low\citep{Fulton2021,Vigan2021}, the probability of a chance-aligned background object is much lower. A background contaminant is further constrained based on archival data. The extremely high proper motion of Eps~Ind~A (4708.15$\pm$0.13mas/year\citep{gaia_dr3}) means that the background location is unobstructed by the star in sufficiently old data. We studied previous observations of Eps~Ind~A, and no stationary background contaminant consistent with the point source is detected. A particularly strong constraint comes from archival Spitzer/IRAC observations\cite{Marengo2006}, which rules out companions brighter than 16.0mag at 8\micron; the point source is 13.16mag at 10.65\micron. Further, the source is unresolved; most galaxies of this brightness would be spatially resolved at MIRI wavelengths. It is unlikely that a stationary background object would have evaded detection in all archival observations while also reproducing the candidate planet properties. A solar system source (e.g. an asteroid) can be excluded since it should move between the F1065C and F1550C observations. A transient source (e.g. a burst) cannot be excluded based on archival data alone, but is strongly  disfavored statistically.

\begin{figure}
    \centering
    \includegraphics[width=\textwidth]{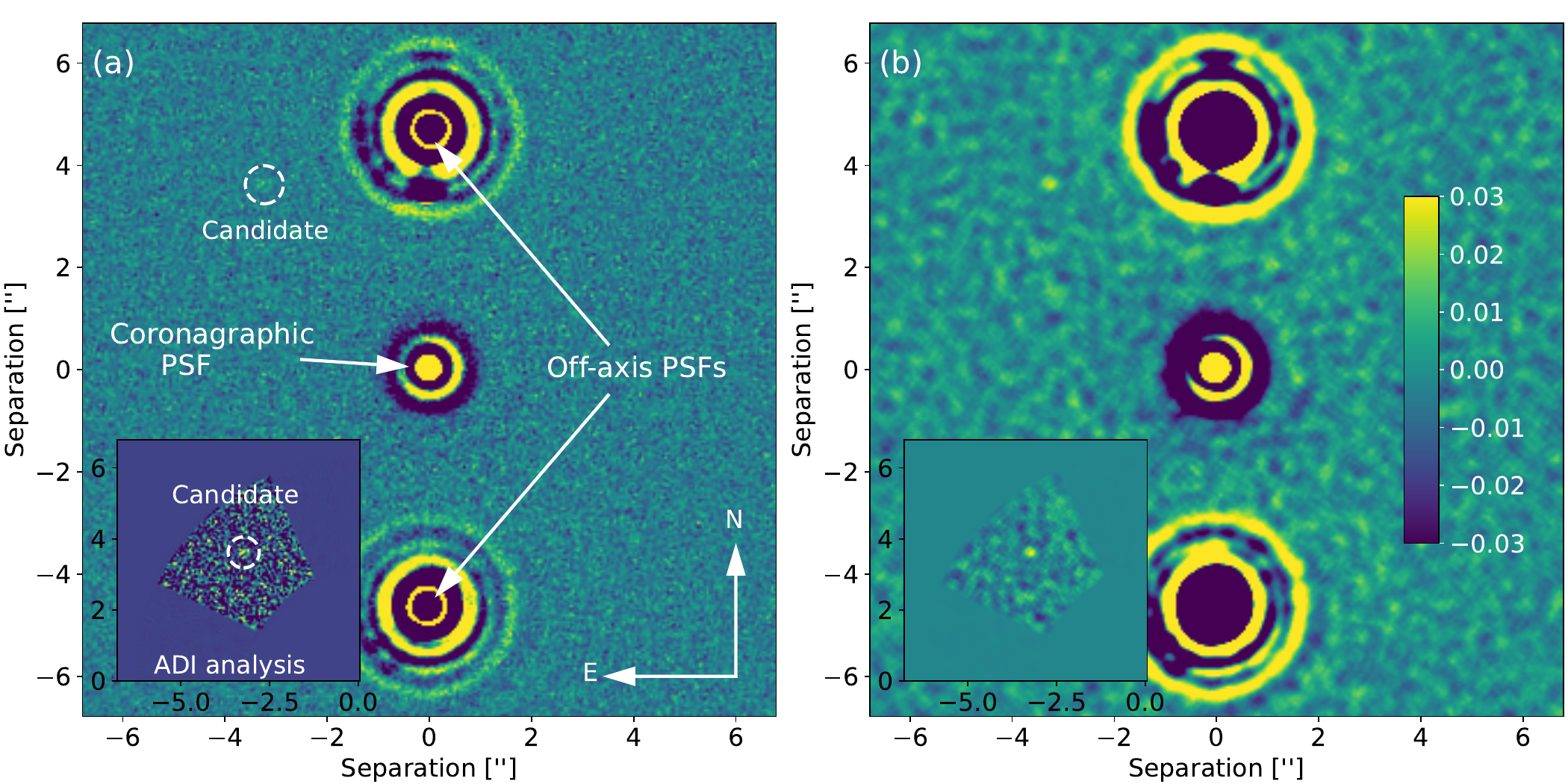}
    \caption{\textbf{A point source is detected at a consistent location in VISIR/NEAR images of Eps~Ind~A.} We show the coadded images, alongside and a principal component analysis on a small patch around the expected target location (inset). Panel (b) shows images (a) after convolution with a tophat function.}
    \label{fig:visir-near}
\end{figure}

We reanalysed archival data collected with VISIR/NEAR\citep{Pathak2021,Viswanath2021}. These coronagraphic observations of Eps~Ind~A were collected over three nights during September 2019, using a broadband filter spanning $\sim$10.-12.5\micron. At $\sim$15au, a planet would have an orbital period of at least several decades, corresponding to little orbital motion in the $\sim$3.8 years between VISIR and JWST observations, while a background source would be shifted by 18'' over this time baseline. Figure \ref{fig:visir-near} shows the VISIR/NEAR observations, where a faint point source at signal-to-noise $\sim$3 is observed, and is consistent with the expected flux and location of the JWST point source. We conclude that this is an archival re-detection of the same object, confirming common proper motion and conclusively demonstrating that it is a planet (hereafter Eps~Ind~Ab).

\begin{figure}
    \centering
    \frame{\includegraphics[width=\textwidth, trim=2cm 0.5cm 0.8cm 1.5cm, clip]{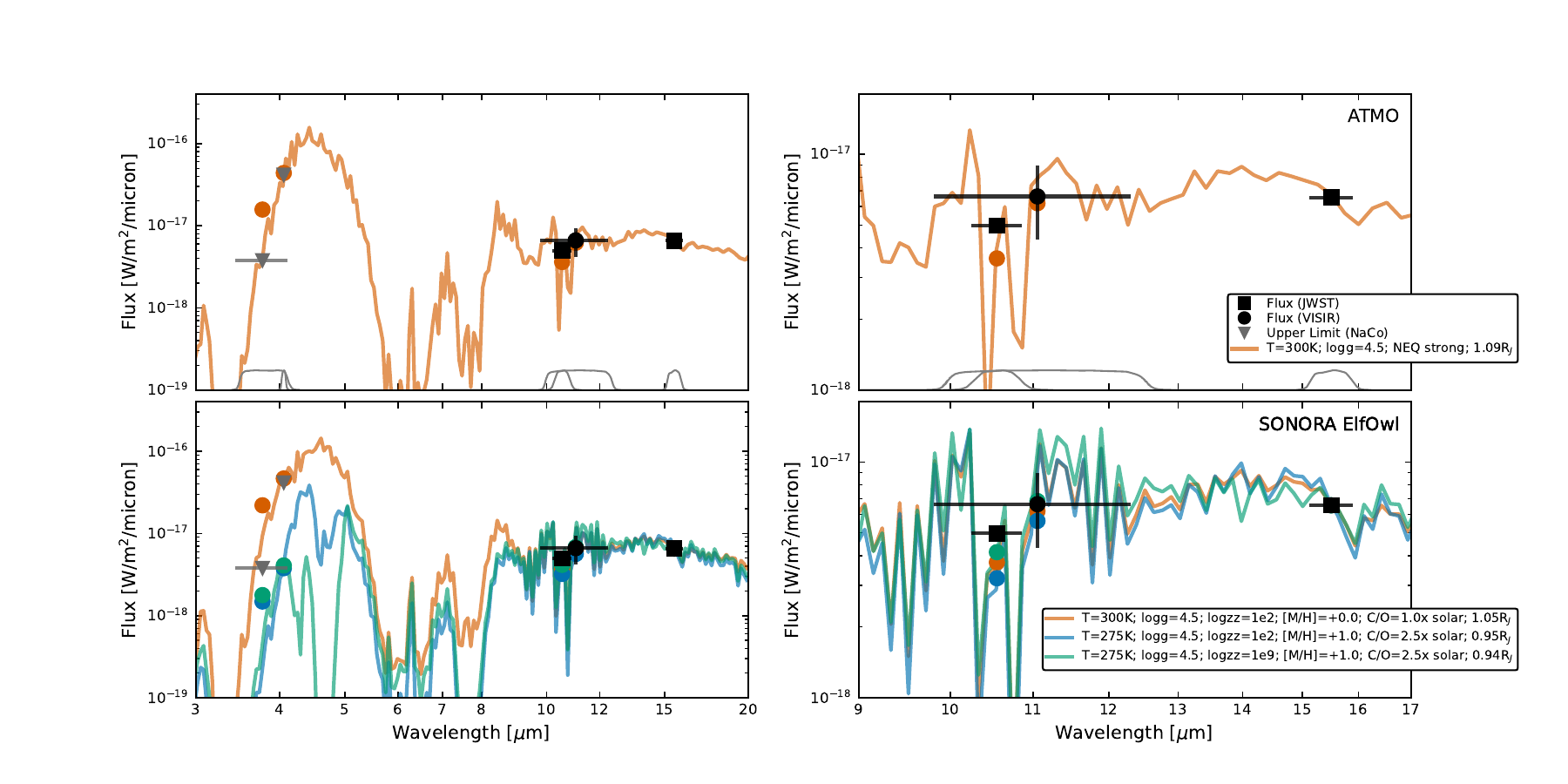}}
    \caption{\textbf{Eps~Ind~Ab is consistent with theoretical atmosphere models with suppressed 3.5-5\micron\ flux.} We show measured photometry (squares, circles, uncertainties are 1$\sigma$) and 5$\sigma$ upper limits (triangles), compared to ``out-of-the-box" model spectra\citep{Phillips2020, Mukherjee2024} (colored lines; circles indicate the integrated flux in each observed filter). Normalized filter profiles are overlayed with grey lines; planet radii are scaled to match the measured 15.50\micron\ photometry. High metallicity and carbon-to-oxygen ratio are required to suppress the 3.5-5\micron\ flux to below the observed upper limit.}
    \label{fig:planetspectra}
\end{figure}

The measured mid-IR photometry (10.65\micron\ and 15.50\micron) of Eps~Ind~Ab is consistent with atmosphere models for a $\sim$300K planet, as demonstrated in Figure \ref{fig:planetspectra}. However, these models typically predict significant emission between 3.5-5\micron, where archival {VLT/NaCo} observations\citep{Janson2009,Viswanath2021} failed to detect Eps~Ind~Ab. Atmospheric models predict a ``window'' for flux at these wavelengths, between absorption features from molecular species including CH$_4$ and H$_2$O\citep{Burrows1997}. This is observed for cold brown dwarfs\citep{Cushing2011,Luhman2024} and expected for cold, giant planets -- in tension with NaCo upper limits (highlighted in Figure \ref{fig:planetspectra}). The NaCo non-detection suggests an additional opacity source in the 3.5-5\micron\ region, through molecular absorption or significant \mbox{cloud/haze} high in the atmosphere. 
While some model grids explore cloudy scenarios \citep{Morley2014,LacyBurrows2023}, these models require patchy clouds and the 3.5-5\micron\ flux is dominated by the cloud-free regions of the atmosphere. This in turn leads to a similarly pronounced emission peak as in the ATMO models (Figure \ref{fig:planetspectra}). Sonora ElfOwl models\citep{Mukherjee2024} with temperature 275K and with high metallicity ([M/H]=1.0), high carbon-to-oxygen ratio (2.5$\times$solar) and strong disequilibrium chemistry\citep{Fegley1996} (high $\log{K_{zz}}$, where $K_{zz}$ is the eddy diffusion coefficient) are compatible with all in-hand observational constraints. This model includes significant CH$_4$, CO$_2$ and CO absorption that suppresses the 3.5-5\micron\ flux, and a modest ammonia absorption feature that reproduces the 10.65\micron\ flux. A high metallicity is somewhat surprising for a super-Jupiter planet; the C/O ratio is consistent with predictions for a planet formed by core accretion beyond the CO$_2$ ice-line\citep{Oberg2011}. Additional photometric and spectroscopic characterization of the planet is crucial to confirm these preliminary indications of the atmospheric C/O ratio, and confirm whether chemistry or haze particles are the true cause of the suppressed 3.5-5\micron\ flux, allowing for a comparison to formation models.

To constrain the dynamical mass of Eps~Ind~Ab, we fitted a single-planet orbit to all published RV data (493 points from 4 facilities over 29 years), as well as the two imaging epochs and the Hipparcos-Gaia astrometry. This gives mass ${6.31}_{-0.56}^{+0.60}$\mj\ and semi-major axis ${28.4}_{-7.2}^{+10}$au. The orbit is eccentric (e=$0.40^{+0.15}_{-0.18}$) and observed at apastron, and is able to explain all in-hand data. While previous studies claimed the presence of a 3.0$\pm$0.1\mj\ planet with semi-major axis 8.8$^{+0.2}_{-0.1}$au\citep{Feng2019,Feng2023,Philipot2023} based on the RV data and astrometric information, the imaged planet imparts a significant acceleration on the host star, and so it is clear that there is no inner exoplanet with the previously published parameters. While the data does not exclude the possibility of second planet in the system, Eps~Ind~Ab is the dominant accelerator of the star. Future works should further explore why the previous RV analyses led to underestimated mass and semi-major axis constraints for this system; this could occur due to over-fitting, or if there are in-fact multiple planets in the system.

\begin{figure*}
    \centering
    \includegraphics[width=\textwidth]{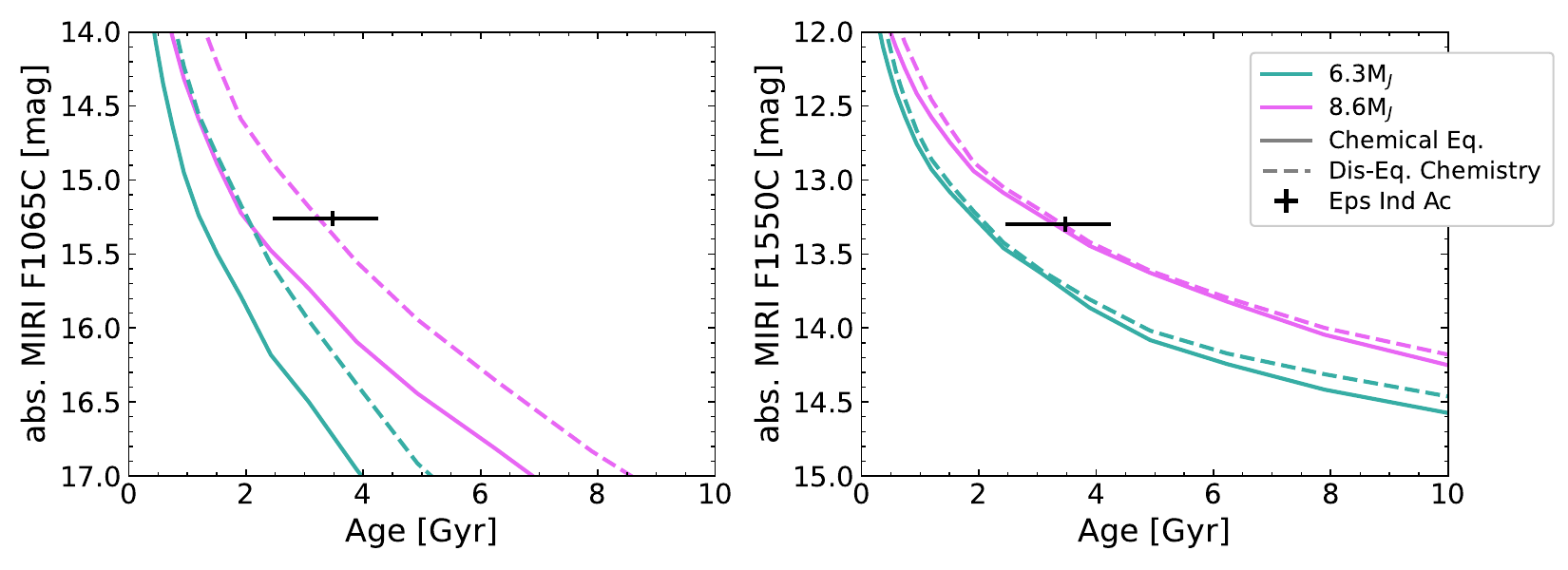}
    \caption{\textbf{Comparison of Eps~Ind~Ab to evolutionary models.} Here we show cooling curves for a 6.3\mj\ and a 8.6\mj\ planet, for models with equilibrium and with non-equilibrium chemistry\citep{Phillips2020}. The mid-IR companion photometry is consistent with a more massive planets, though these models do not incorporate the high metallicity and carbon-to-oxygen ratio of the planet, and the correspondingly low 3.5-5\micron\ flux.}
    \label{fig:evo_models}
\end{figure*}

Evolutionary models predict a temperature of $\sim$280K and a luminosity of $\log(L/L_\odot)\sim-7.2$ for a 3.5~Gyr, 6.3\mj\ planet\citep{Phillips2020}. In Figure \ref{fig:evo_models} we present evolutionary models\citep{Phillips2020} for two planets: a 6.3\mj\ object (matching the dynamical mass) and a 8.6\mj\ planet (matching the observed photometry), considering both equilibrium and disequilibrium chemistry. Based only on the age and MIRI photometry for Eps~Ind~Ab, the planet is overluminous relative to models. However, the extent of the flux suppression at 3.5-5\micron\ remains unclear, and this impacts the derived temperature of the planet. The best-fit model in Figure \ref{fig:planetspectra} incorporates this suppressed flux, and has a temperature of 275K, very similar to predictions from evolutionary models. Additional photometry between 3.5-5\micron\ is crucial to confirm the extent of flux suppression, better constrain the temperature of Eps~Ind~Ab, and thereby allow a more thorough comparison to evolutionary models.

Eps~Ind~Ab is the coldest exoplanet to be directly imaged, with a similar temperature to the coldest field brown dwarf (WISE 0855)\citep{Luhman2024}. The system is also co-moving with a widely-separated brown dwarf binary\citep{Scholz2003,McCaughrean2004}, making it a particularly valuable laboratory for comparative studies of substellar objects with a shared age and formation location. The exoplanet detection highlights the power of using indirect evidence to target direct detection efforts. Even though the detected planet does not match the previously claimed exoplanet properties, long-term RV information provided a clear signpost for the value of imaging this target. The photometry of Eps~Ind~Ab suggests high metallicity and carbon-to-oxygen ratio, and future works should aim to measure the planet flux between 3.5-5\micron\ to determine the extent of the flux suppression. Future works should also aim to expand the sample of cold exoplanets, and determine whether other cold planets show the same low 3.5-5\micron\ flux, indicating high metallicity and high carbon-to-oxygen ratio; two candidate companions to white dwarf stars\citep{Mullally2024} and several upcoming JWST observations of nearby accelerating stars\citep{jwst_3337,jwst_5229} may allow for a small sample of cold, solar-age exoplanets to be assembled. The impact of atmospheric absorption is also critical to consider when deriving mass detection limits, which are used to calculate occurrence rates, and designing future observations of cold exoplanets. The bright flux and wide separation of Eps~Ind~Ab mean the planet is ideally suited to spectroscopic characterization efforts, allowing the metallicity and carbon-to-oxygen ratio to be more accurately constrained.

\clearpage

\bmhead{Data Availability}
All observational data in this work are available through public data archives. In particular, the JWST data were collected through the General Oberver (GO) program \#2243 (PI Matthews) and are available through the MAST database (\url{https://mast.stsci.edu}). Spitzer data are collected through programmes 90 (PI Werner) and 41 (PI Rieke) and are available though the Spitzer Heritage Archive (\url{https://irsa.ipac.caltech.edu/applications/Spitzer/SHA/}). VLT/NEAR and VLT/SPHERE data are collected through programmes 60.A-9107/103.201H (PI Meyer) and 110.25BR (PI Matthews) respectivly, and are available through the ESO science archive (\url{http://archive.eso.org/cms.html}). RV data and atmospheric models are drawn from previously published literature, as indicated in the text.

\bmhead{Code Availability}
Data reduction was carried out using the publicly available \texttt{spaceKLIP} package (\url{https://github.com/kammerje/spaceKLIP}). Orbit fitting was carried out using the publicly available \texttt{orvara} package (\url{https://github.com/t-brandt/orvara}). We used various functions from \texttt{astropy}, \texttt{pandas}, \texttt{matplotlib} and \texttt{seaborn} to carry out analysis and create figures.

\bmhead{Author Contributions}
ECM designed the science program, performed data analysis of JWST/MIRI, VLT/SPHERE, Spitzer/IRAC and Spitzer/MIPS data, carried out statistical analysis of the background and companion hypotheses, contributed to orbit fitting, carried out comparison to atmospheric and evoutionary models, reanalysed the stellar spectrum, led the interpretation of these results, and wrote the manuscript. AC designed the MIRI observing sequence and supported the JWST/MIRI data reduction. PP carried out the VISIR/NEAR data reduction with input from MK. CVM and MWP provided theoretical interpretation of the photometric constraints and upper limits. SKPM, ED, JH and CLK provided an alternative JWST/MIRI data reduction. FF, JB, AML, FP contributed to orbit fitting and dynamical mass estimates for the target. MJB provided a reanalysis of the VLT/NACO data. LAB provided interpretation of possible galactic contaminants. IJMC and ThH contributed to interpretation of the results. DP contributed to ground-based vetting of the reference target. The work is based on a JWST proposal written by ECM with contributions from AC, FF, MWP, CVM, IJMC, ED and JB. All authors provided discussion of the results and commented on the manuscript.

\bmhead{Competing Interests}
The authors declare no competing financial interests.

\bmhead{Correspondence}
Correspondence and requests for materials should be addressed to E.~C.~Matthews.

\bmhead{Acknowledgements}

We thank Markus Janson, Jackie Faherty and Tim Brandt for discussions that aided the preparation of this paper.

This work is based on observations made with the NASA/ESA/CSA James Webb Space Telescope. The data were obtained from the Mikulski Archive for Space Telescopes at the Space Telescope Science Institute, which is operated by the Association of Universities for Research in Astronomy, Inc., under NASA contract NAS 5-03127 for JWST. These observations are associated with program \#2243.

This work makes use of observations collected at the European Southern Observatory under ESO programmes 110.25BR and 60.A-9107 (103.201H).
The NEAR project was made possible through contributions from the Breakthrough Foundation and Breakthrough Watch program, as well as through contributions from the European Southern Observatory and the NEAR consortium.

This work makes use of observations made with the Spitzer Space Telescope, which was operated by the Jet Propulsion Laboratory, California Institute of Technology under a contract with NASA.

Support for program \#2243 was provided by NASA through a grant from the Space Telescope Science Institute, which is operated by the Association of Universities for Research in Astronomy, Inc., under NASA contract NAS 5-03127.
Portions of this research employed High Performance Computing (HPC) resources supported by the University of Arizona TRIF, UITS, and Research, Innovation, and Impact (RII) and maintained by the UArizona Research Technologies department.
Part of this research was carried out at the Jet Propulsion Laboratory, California Institute of Technology, under a contract with the National Aeronautics and Space Administration (NASA).
The contributions of D.P have been carried out within the framework of the NCCR PlanetS supported by the Swiss National Science Foundation under grants 51NF40\_182901 and 51NF40\_205606.
This project has received funding from the European Research Council (ERC) under the European Union's Horizon 2020 research and innovation programme (COBREX; grant agreement n° 885593). AML and FP received funding by PSL/OCAV.

\section*{Methods}

\setcounter{figure}{0}
\setcounter{table}{0}
\renewcommand{\figurename}{Extended Data Figure}
\renewcommand{\tablename}{Extended Data Table}
\renewcommand{\thesubsection}{\arabic{subsection}}

\subsection{JWST observations \& data reduction}

We observed Eps~Ind~A with the JWST/MIRI coronagraphic imager \citep{Rieke2015,Boccaletti2015,Boccaletti2022} on 2023 Jul 03 (JWST GO program 2243\citep{Matthews_jwstgo2243}) in a single sequence. Our observations consisted of science, reference, and background images, detailed in Extended Data Table \ref{tab:jwstobservations}. We collected images with two narrow-band coronagraphic filters, F1065C and F1550C, integrating for 3772s and 3922s respectively in the two filters. Each coronagraphic filter has a dedicated 4-quadrant phase-mask coronagraph (4QPM)\citep{Rouan2000} for suppression of the stellar PSF. Dedicated background observations of an empty patch of sky were collected at the start of the sequence, to allow subtraction of the ``glow stick" stray-light feature, first identified in MIRI commissioning \citep{Boccaletti2022}. For each filter and each of the science and reference targets, we collected background observations using the same integration as a single science/reference dither, and repeated this process over two dither positions. We used the same empty field for all background observations.

\begin{table*}[ht]
    \footnotesize
    \centering
    \begin{tabular}{llllllll}
        \toprule
        Target & Telescope Coordinates & Filter & Readout & Groups/Int & Ints/Exp & Dithers & Total Time [s] \\
        \midrule
        \midrule
        background & 22:17:54.000 -57:28:50.00 & F1550C & FASTR1 & 811 & 14 & 2 & 5448.885 \\ 
        background & 22:17:54.000 -57:28:50.00 & F1550C & FASTR1 & 1168 & 14 & 2 & 7844.726 \\ 
        background & 22:17:54.000  -57:28:50.00 & F1065C & FASTR1 & 89 & 122 & 2 & 5262.893 \\ 
        background & 22:17:54.000  -57:28:50.00 & F1065C & FASTR1 & 128 & 122 & 2 & 7543.688 \\ 
        \midrule
        DI Tuc & 22:16:08.195  -57:34:05.06 & F1065C & FASTR1 & 89 & 122 & 5 & 13157.234 \\ 
        DI Tuc & 22:16:08.195  -57:34:05.06 & F1550C & FASTR1 & 811 & 14 & 5 & 13622.213 \\ 
        \midrule
        Eps~Ind~A & 22:03:33.001  -56:48:09.13 & F1550C & FASTR1 & 1168 & 14 & 1 & 3922.363 \\ 
        Eps~Ind~A & 22:03:33.001  -56:48:09.13 & F1065C & FASTR1 & 128 & 122 & 1 & 3771.844 \\ 
        \botrule
    \end{tabular}
    \caption{JWST/MIRI observations for Eps~Ind~A. All observations were carried out sequentially on Jul 3, 2023, and have one exposure per dither position.}
    \label{tab:jwstobservations}
\end{table*}

\begin{figure}
    \centering
    \includegraphics[width=\textwidth]{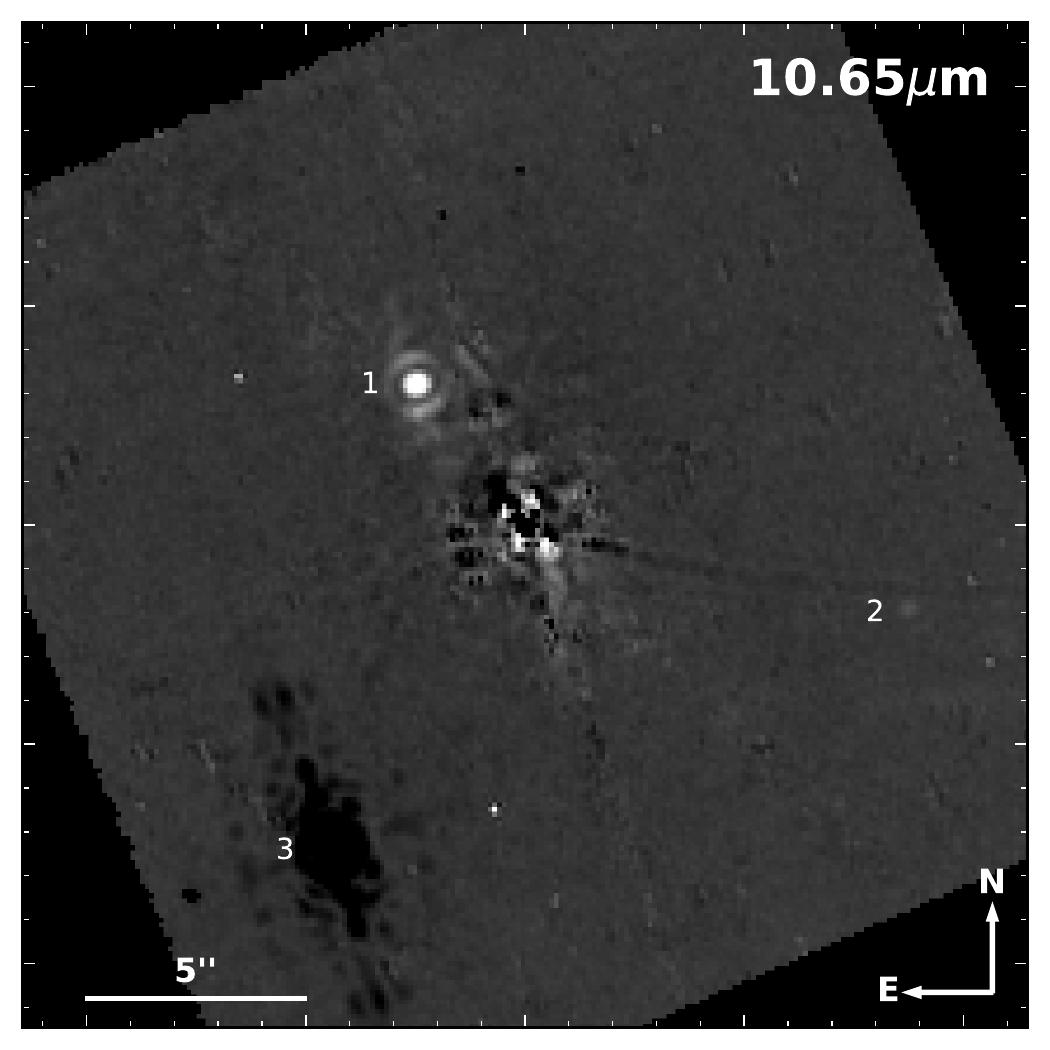}
    \caption{Full field-of-view JWST/MIRI coronagraphic images of Eps~Ind~A in the 10.65\micron\ filter. Two visual companions to Eps~Ind~A, and one visual companion to DI Tuc, are identified: (1) is Eps~Ind~Ab, (2) is the background star Gaia DR3 6412595290591430784 (also identified in the Spitzer 8\micron\ observations of the background field) and (3) is the background star Gaia DR3 6411654761473726464 in the field of the DI Tuc observations.}
    \label{fig:jwstdata_full}
\end{figure}

We collected PSF reference images of the star DI Tuc (HD211055) to allow for reference differential imaging (RDI)\citep{Smith1984}. We collected 5 reference images for each filter, using the 5-point small grid dither technique \citep{Soummer2013,Lajoie2016} to account for the imperfect pointing of the JWST (commissioning data suggest a pointing stability of $\sim$5-10mas\citep{Rigby2023}). Integration times per dither position were 2631s and 2724s for the F1065C and F1550C observations respectively, with these integrations chosen to match the peak flux of the Eps~Ind~A observations at each dither position. In advance of the JWST observations, we also screened the reference star for faint companions, to ensure it was suitable for differential imaging. We observed DI Tuc with VLT/SPHERE on 2023-04-13 (program 110.25BR, PI Matthews) and collected 140 images, each with DIT=0.837s (total exposure time 117s), in the broadband H filter ($\lambda$=1.6255\micron, $\Delta\lambda$=0.291\micron). Images were dark- and flat-fielded and bad pixels corrected, and we then aligned all images based on the peak flux position, and coadded across the stack of observations. No companions were detected anywhere in the SPHERE field of view ($\sim12''\times12''$). Two companions are in the \textit{Gaia} catalog and the JWST field of view, but outside the SPHERE field of view (separations 6.5'' and 8.2''): these are sufficiently widely separated that they do not impact the PSF subtraction, though the brighter of these is clearly visible in the reduced data (Extended Data Figure~\ref{fig:jwstdata_full}).

We used the \texttt{spaceKLIP}\footnote{\url{https://github.com/kammerje/spaceKLIP}} pipeline \citep{Kammerer2022,Carter2023} for the JWST data analysis. This pipeline provides various high-contrast imaging specific functionalities for JWST data, and in particular makes use of the \texttt{jwst}\footnote{\url{https://jwst-pipeline.readthedocs.io/en/latest/}} pipeline \citep{Bushouse2024} and the \texttt{pyKLIP}\footnote\url{{https://pyklip.readthedocs.io/en/latest/}} python package \citep{Wang2015} for several key reduction steps. Our reduction follows the steps outlined in detail in \citep{Carter2023}, and is briefly summarized here. We started from the Stage 0 (\texttt{*uncal.fits)} files as generated by the \texttt{jwst} pipeline. We performed ramp-fitting, calibrated the images from detector units to physical units, flat-fielded, and subtracted the background images. We then subtracted the stellar PSF using reference differential imaging. We explored various parameters; the reductions shown in this work have PCA with a single subsection, a single annulus, and 8 PCA modes removed.

\begin{figure}
    \centering
    \includegraphics[width=0.7\textwidth]{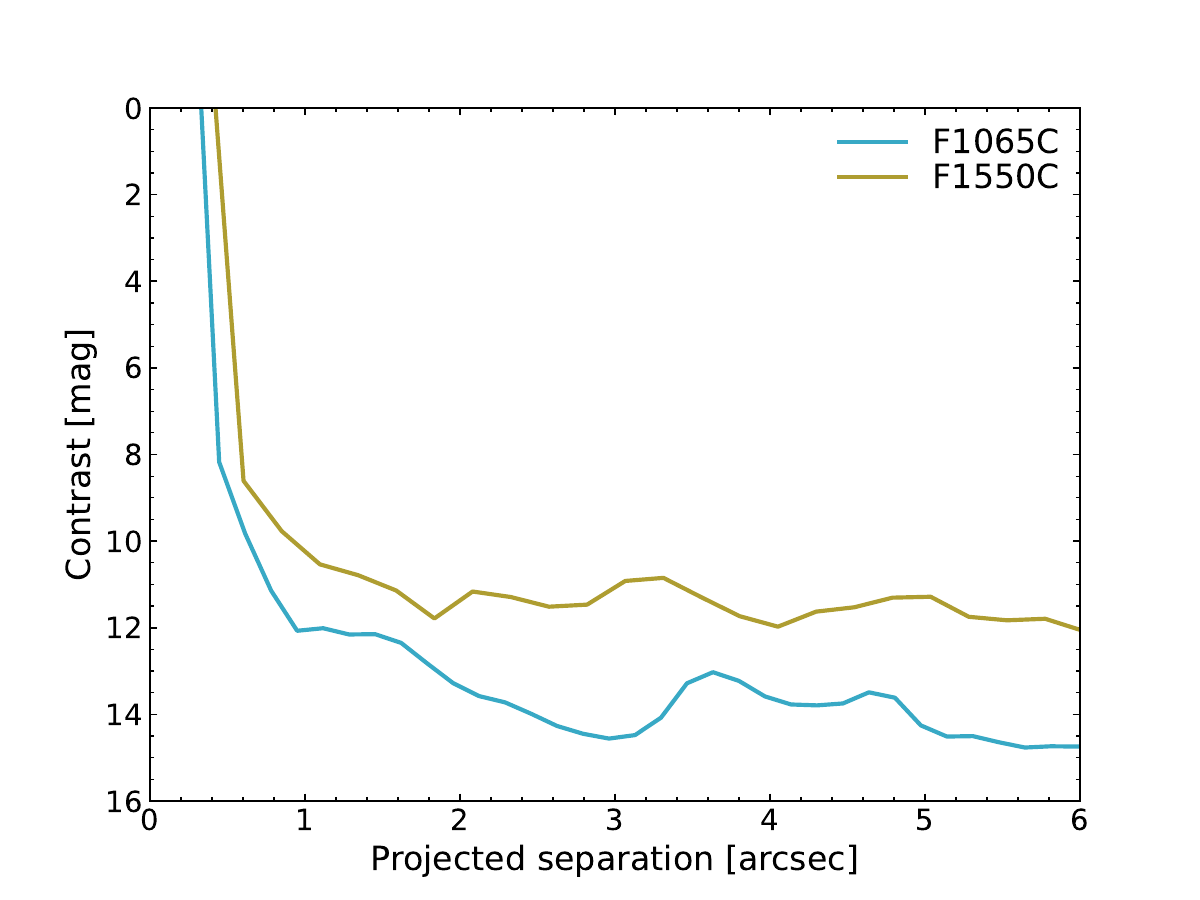}
    \caption{Contrast sensitivity of the MIRI observations. Curves are 5$\sigma$ sensitivities calculated with spaceKLIP for each coronagraphic filter.}
    \label{fig:contrast}
\end{figure}

This process provides excellent starlight suppression for the F1550C images, with the contrast performance provided in Extended Data Figure \ref{fig:contrast}. The stellar PSF is less well matched between the science and reference images for F1065C, likely due to the larger time baseline between these observations (the order of observations is F1065 reference, F1550C reference, F1550C science, F1065C science, as listed in Extended Data Table \ref{tab:jwstobservations}). This reduces the planet sensitivity at close separations, and some residual starlight is clearly visible in these images. The full field of view for the 10.65\micron\ images is shown in Extended Data Figure \ref{fig:jwstdata_full}, and the central portion of both images is included in Figure \ref{fig:jwstdata}. Three astrophysical point sources are detected at F1065C:
\begin{enumerate}
\item A bright source to the NE of Eps~Ind~A, which was not previously known. This source is confirmed to be the exoplanet Eps~Ind~Ab in the current work.
\item A faint source to the W of Eps~Ind~A is a background star. This source corresponds to Gaia DR3 6412595290591430784, and is also detected in archival Spitzer/IRAC observations.
\item A negative source to the SE of Eps~Ind~A is a widely separated companion to DI Tuc, that is subtracted in our RDI reduction. This source corresponds to Gaia DR3 6411654761473726464.
\end{enumerate}

We determined the companion properties using the forward-model PSF fitting routine provided by \texttt{pyKLIP} and implemented in \texttt{spaceKLIP}. We generated a PSF template at the location of the companion using \texttt{webbpsf\_ext}\footnote{\url{https://github.com/JarronL/webbpsf_ext}}. We then fitted the location and intensity of this PSF to the data to measure the photometry and astrometry of the observations; these values are listed in Extended Data Table \ref{tab:phot_ast}.

We also reduced the data using a non-negative matrix factorization (NMF) approach \citep{Ren2018,PM2023} to verify our results. NMF involves decomposing an image matrix into two nonnegative matrices; one extracts features and the other adds weights to reconstruct the original matrix. The features matrix extracted from the RDI PSF library is used to model the stellar PSF in the target image frames, and then PSF subtraction is performed to image the exoplanet. We performed the NMF reduction using all frames from the DI Tuc PSF library to construct components. For F1065C, this involved using 605 frames to construct 605 components, and for F1550C the number of frames and components was 65. We tried alternative approaches, including using the entire PSF library with fewer components and using a smaller PSF library with components selected using their closeness to the target frames (using the Euclidean norm), but using the entire library with the maximum number of components consistently produced the best images. 

\subsection{Background constraints derived from archival data products}
\label{methods:backgroundobs}

Analysis of archival data provides tight constraints on a possible background contaminant. If the point source were an extragalactic contaminant, it would have no proper motion: it should be found 4.1'' to the north-east of the JWST Eps~Ind~A position, namely at RA=22:03:33.17 and Dec=-56:48:06.0. For a background object, these absolute RA and Dec coordinates should be unchanged as Eps~Ind~A moves across the sky at a rate of 4.7arcsec/year, and would be well resolved from the star in sufficiently old archival data (allowing for a sufficient movement of the foreground star, relative to the background position). Even if a background contaminant were a distant object within the Milky Way, it would have a proper motion at most a few mas per year, and at most a few tens of mas over the $\sim$2 decades spanned by data products presented here. We searched the Gaia\citep{gaia_dr3} and 2MASS\citep{Skrutskie2006} catalogs, and reanalysed Spitzer/IRAC and Spitzer/MIPS data, and did not identify any background sources consistent with the point source.

\begin{figure}
    \centering
    \includegraphics[width=0.86\textwidth]{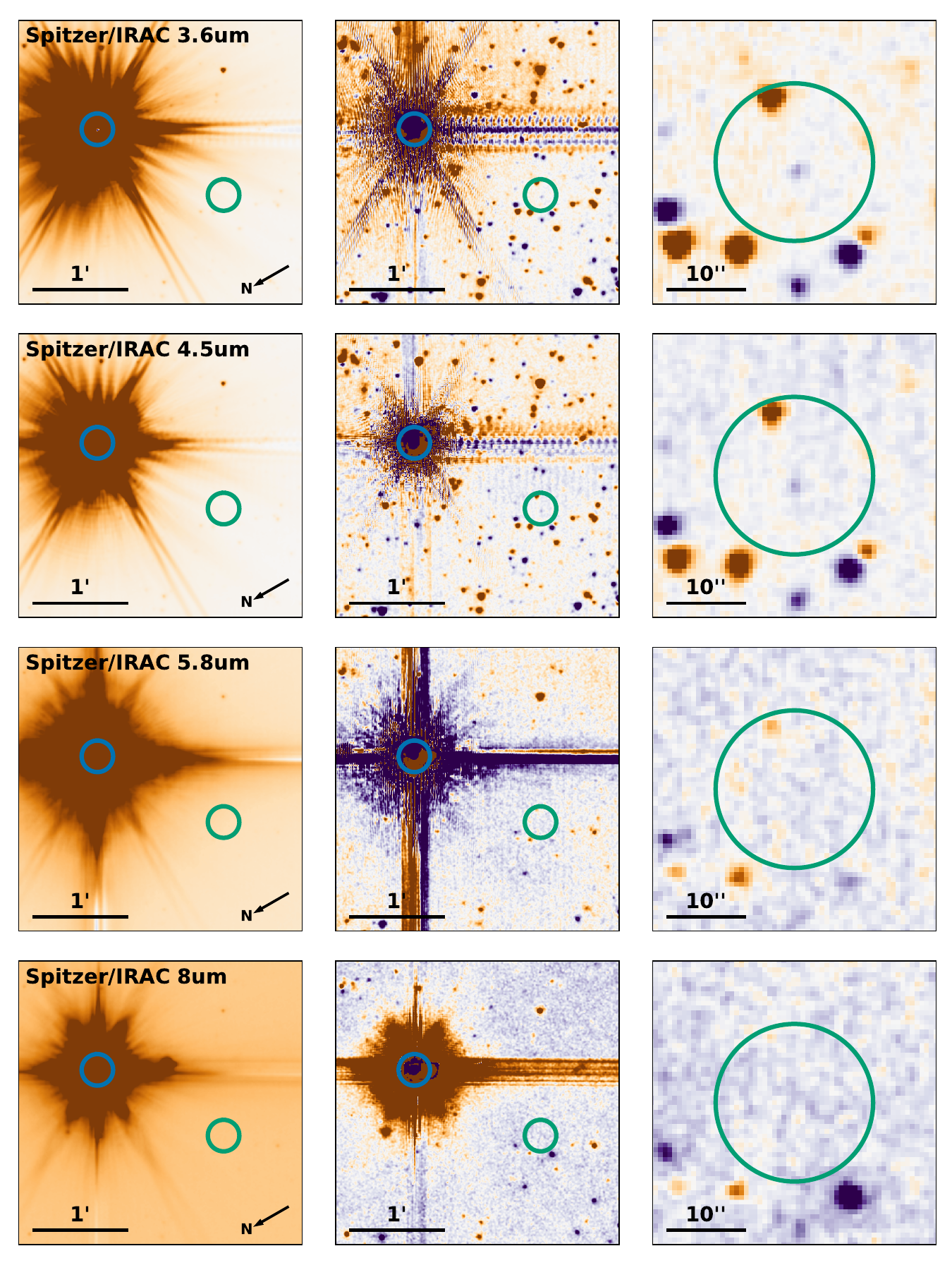}
    \caption{Spitzer IRAC images of Eps~Ind~A. The three columns show the science-ready Eps~Ind~A image from the Spitzer archive (left), an RDI processed image with Eps Eri subtracted from Eps~Ind~A (center), and the RDI processed image zoomed at the JWST epoch position of Eps~Ind~A (right). For the RDI processed image, purple (negative) sources are in the field of the reference star, and orange (positive) sources are in the field of Eps~Ind~A. Blue and green circles have radii 10'', and are centered on the Spitzer \& JWST epoch positions of Eps~Ind~A respectively. If Eps~Ind~Ab were a background object, it would be $\sim$4'' from the JWST epoch position of Eps~Ind~A. This is well within the blue circle. One source is identified within this circle at 3.6\micron, 4.5\micron\ and marginally at 5.8\micron: this is a background star also identified in the JWST/MIRI images (see text for details).}
    \label{fig:spitzer-irac}
\end{figure}

Spitzer observations provide excellent background constraints, and bracket the wavelengths of our JWST observations. Several Spitzer observations of Eps~Ind~A were carried out during 2004, and Eps~Ind~A subsequently moved $\sim$90'' in the 19 years between the Spitzer and JWST observations. Eps~Ind~A was observed with Spitzer/IRAC on 2004-05-01 (program ID 90; PI Werner), with data collected in all four IRAC channels (nominally 3.6\micron, 4.5\micron, 5.8\micron, 8.0\micron). The star was used primarily as a PSF reference star for a program studying the Eps Eri debris disk \citep{Marengo2006,Marengo2009,Backman2009}. We reanalyzed these data, starting from the science-ready mosaic files from the Spitzer Heritage Archive\footnote{\url{https://irsa.ipac.caltech.edu/applications/Spitzer/SHA/}} for each of the four wavelength channels. Even though the background position is well separated from the star at this epoch, there is still significant signal in the PSF wings (Extended Data Figure~\ref{fig:spitzer-irac}). We used the Eps Eri observations from the same program as a PSF reference to perform reference differential imaging (RDI) and subtract the stellar PSF wings, thereby allowing the best possible contrast at the background location. We used a grid-based approach to optimize the spatial (x, y) location and flux scaling of the Eps Eri images, and thereby provide the best match to the PSF wing structure of the Eps~Ind images, and then subtracted this best-match image from the Eps~Ind data to produce the RDI-subtracted images. The science-ready images and RDI-subtracted images are shown in Extended Data Figure \ref{fig:spitzer-irac}, with the position of Eps~Ind~A during the Spitzer and JWST epochs highlighted. We calculated the flux limits based on the pixel variance in the RDI-subtracted image, with known sources masked, and validated our limits by performing aperture photometry on identified sources in the image. 

\begin{figure}
    \centering
    \includegraphics[width=\textwidth]{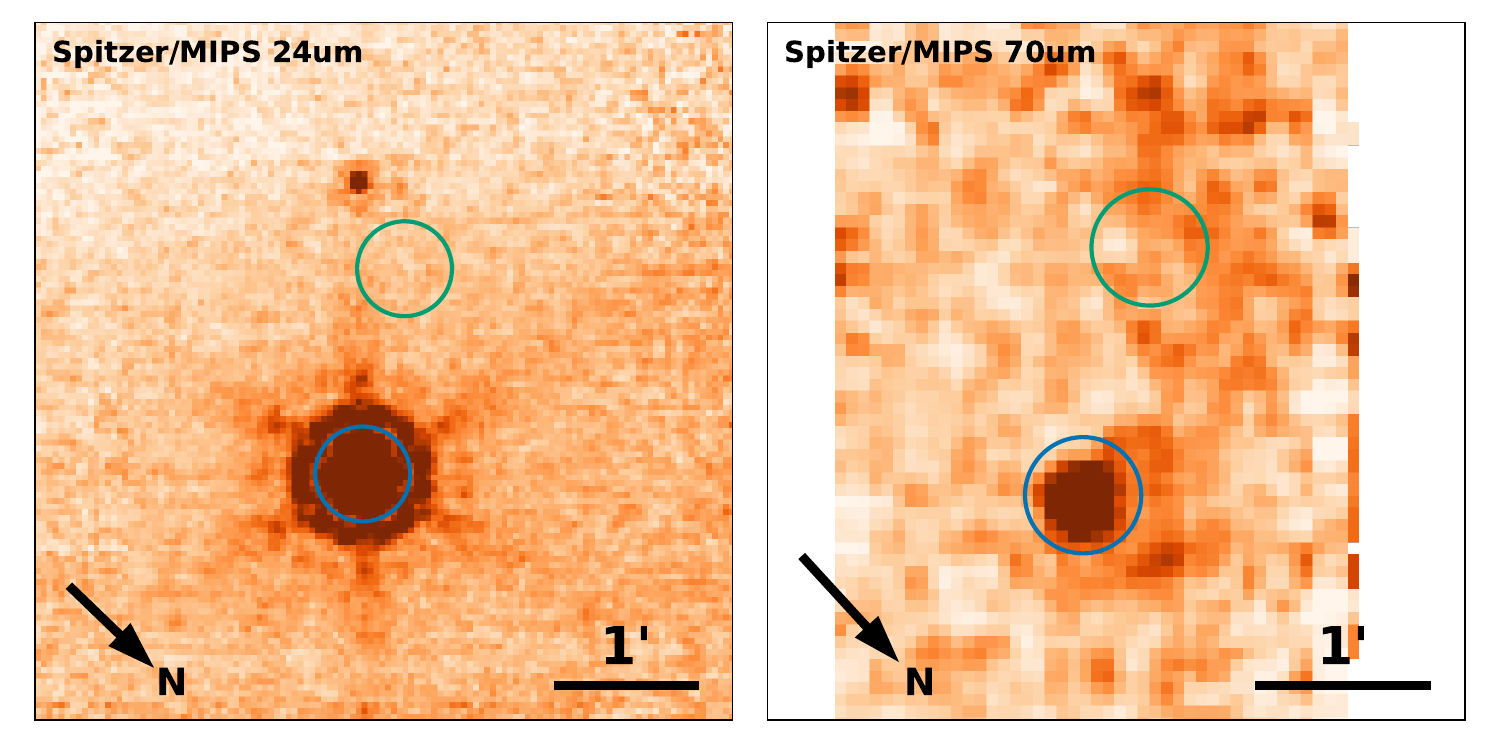}
    \caption{Spitzer MIPS 24\micron\ and 70\micron\ images of Eps~Ind~A. Blue and green circles have radii 20'', and are centered on the Spitzer \& JWST epoch positions of Eps~Ind~A respectively; no sources are detected at the background position (the JWST epoch position of Eps~Ind~A). The 70\micron\ data covers only a narrow strip around the Eps~Ind~A location, and white regions at the edge of the box indicate regions outside the telescope field of view.}
    \label{fig:spitzer-mips}
\end{figure}

Eps~Ind~A was also observed with Spitzer/MIPS on 2004-10-13 (program ID 41; PI Rieke)\cite{Gaspar2013}, with data collected in all three MIPS channels (nominally 24\micron, 70\micron, and 160\micron). We analysed the science-ready mosaic files from the Spitzer Heritage Archive. Eps~Ind~A is unsaturated in the MIPS images, and the JWST position is well resolved from the star (Extended Data Figure \ref{fig:spitzer-mips}) and within the field of view for the 24\micron\ and 70\micron\ channels. We calculated the flux upper limit based on the standard deviation of pixel values in a large box around the position of Eps~Ind~A, excluding pixels within 1 arcmin of the star. No sources are observed within 10'' of the JWST Eps~Ind~A position in the Spitzer IRAC 8.0\micron\ images, or in any of the MIPS images. One source is identified $\sim$9'' west of the Eps~Ind~A position in the 3.6\micron, 4.5\micron\ and 5.8\micron\ images. This source is also identified in the JWST/MIRI images, and is labeled \textit{2} in Extended Data Figure \ref{fig:jwstdata_full}. The location and magnitude measurements correspond to that of the star Gaia DR3 6412595290591430784: this is a background object, not associated with Eps~Ind~A.

\begin{figure}
    \centering
    \includegraphics[width=0.7\textwidth]{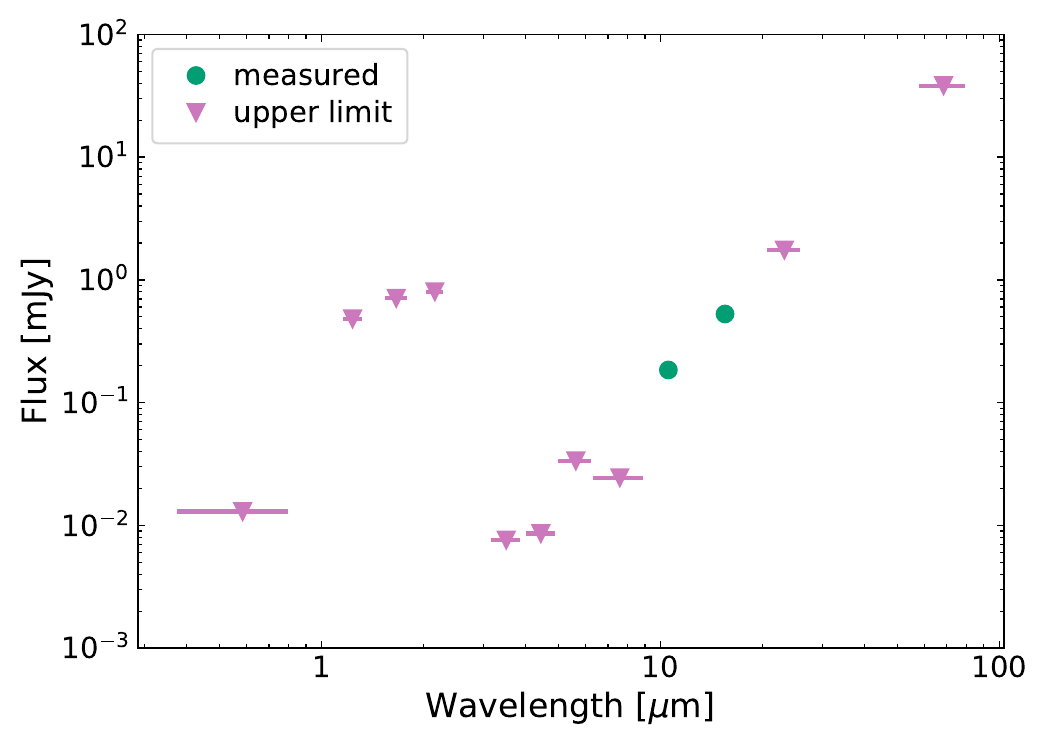}
    \caption{Spectrum (measured flux \& upper limits) of the point source if it were a background object. These upper limits apply only if the point source were a stationary background at RA=22:03:33.17 and Dec=-56:48:06.0. Constraints for the planetary case (which require high-contrast facilities to resolve a source that moves with same proper motion as the host star) are given in Figure \ref{fig:planetspectra}. Green circles indicate the measured JWST photometry, while pink triangles indicate upper limits from Gaia, 2MASS, Spitzer/IRAC and Spitzer/MIPS. Most stationary background contaminants would not reproduce the 10.65\micron\ and 15.50\micron\ flux while evading detection at other wavelengths.} 
    \label{fig:backgroundspectrum}
\end{figure}

A background spectrum for the point source is shown in Extended Data Figure \ref{fig:backgroundspectrum}: this indicates the flux and upper limits that would apply to the object if it were a stationary background object, or a background object that moved by less than a few hundred mas/yr. That is, these upper limits do \textit{not} apply to the planet, which shares proper motion with Eps~Ind~A and is not present at this location in the background observations. The 8.0\micron\ images provide a particularly stringent constraint on the background scenario: we rule out background objects to 16.07mag at this wavelength, while the point source is 13.16mag at 10.65\micron. A background object would have to be extremely red to be compatible with both measurements.

\subsection{Re-analysis of archival VISIR/NEAR data} \label{methods:visirnear}

We revisited archival VISIR/NEAR\citep{Lagage2004,Kasper2017} observations of Eps~Ind~A. We started from previously published reduced datacubes \citep{Pathak2021}, for which frames were calibrated and aligned, a frame selection was carried out, and the frames were binned and spatially high-pass filtered. The NEAR sensitivity was optimized for small separation (0.5”-3”) around stars. Ground-based thermal infrared observation require fast chopping to subtract quickly varying sky background and excess low-frequency detector noise. With NEAR, chopping was done using the deformable secondary mirror of the VLT which provides a maximum chop throw of about 4.5 arcseconds. Angular differential imaging reduction relies on the subtraction of images taken at a different field orientation thus leaving residuals of real sources in the azimuthal direction. As the separation of the imaged planet from the host star is similar to the chop throw, these residuals mask the planet if data reduction by ADI were used. To avoid this effect we simply co-added data without performing ADI, since the expected separation is well within the background-limited regime rather than the contrast-limited regime. We also performed a LOCI PCA ADI \citep{Lafreniere2007} analysis with 3 principal components in a small patch around the expected companion location. This further helps to improve the planet signal and suppress the thermal background. In each case, we also convolved the final images with a tophat to further enhance any true signals in the data. Both images are shown in Figure \ref{fig:visir-near}. To further confirm the source, we also looked at the companion position relative to the off-axis PSFs. We centered, aligned, and derotated the frames relative to this position. We found that the planet signal was also present in the final averaged convolved frames relative to the off-axis PSFs, negating the possibility of the source being a speckle or a ghost from internal reflections. We assume an error of half FWHM on the companion position, due to the high-spatial resolution and sampling of the NEAR. The planet is at separation $4.82\pm0.16"$ and at position angle of $40-45^o$ anti-clockwise of North. For flux estimation, we used planet injection and recovery tests. Since the planet signal is thermal background limited, we put a conservative range of $(4-8)\times10^{-5}$ on contrast and a range of $0.18-0.35~$mJy on flux. These values are also included in Extended Data Table \ref{tab:phot_ast}.

\begin{table}[ht]
    \centering
    \footnotesize
    \begin{tabular}{llp{14mm}|ll|ll}
         \toprule
         Instrument & Filter & Date & $\rho$ & $\theta$ & Flux & 5$\sigma$ Limit \\
         & & & [arcsec] & [deg] & [W/m$^2$/\micron] & [W/m$^2$/\micron] \\
         \midrule 
         \midrule
         JWST/MIRI & F1065C & 2023-07-03 & 4.095$\pm$0.010$^{(a)}$ & 38.17$\pm$0.44$^{(a)}$ & $(4.98\pm0.15)\times 10^{-18}$ & -- \\
         \midrule
         JWST/MIRI & F1550C & 2023-07-03 & 4.114$\pm$0.010 & 37.39$\pm$0.43 & $(6.55\pm0.20)\times 10^{-18}$ & -- \\
         \midrule
         VLT/VISIR & NEAR & 2019-09-14 \newline 2019-09-15 \newline 2019-09-17 & $4.82\pm0.164$ & $42.5\pm2.5$ & $(6.6\pm2.2)\times 10^{-18}$ & -- \\ 
         \midrule
         VLT/NaCo & L' & 2018-10-12 \newline 2018-10-26 \newline 2018-11-03 \newline 2018-11-04 & -- & -- & -- & 3.78$\times 10^{-18}$  \\
         \midrule
         VLT/NaCo & NB4.05 & 2008-10-31 \newline 2008-09-02 & -- & -- & -- & 4.16$\times 10^{-17}$ \\
         \botrule
    \end{tabular}
    \caption{Astrometry, Photometry and upper limits for Eps~Ind~Ab. JWST \& VLT/VISIR constraints are from this work; VLT/NaCo constraints are from \cite{Viswanath2021} (L', their ``special LOCI'' reduction with our stellar calibration from Section \ref{methods:stellarphotometry}) and \cite{Janson2009} (NB4.05, we quote a median sensitivity of the observations between 3'' and 8''). Values for multi-epoch visits (VLT/VISIR, VLT/NaCo) correspond to a combined analysis of all visits.\newline(a) Note that for F1065C the star is not well centered behind the coronagraph, and the astrometry should be considered unreliable.}
    \label{tab:phot_ast}
\end{table} 

\subsection{Orbital constraints} \label{methods:orbit}

\begin{figure}
    \centering
    \includegraphics[width=0.37\textwidth]{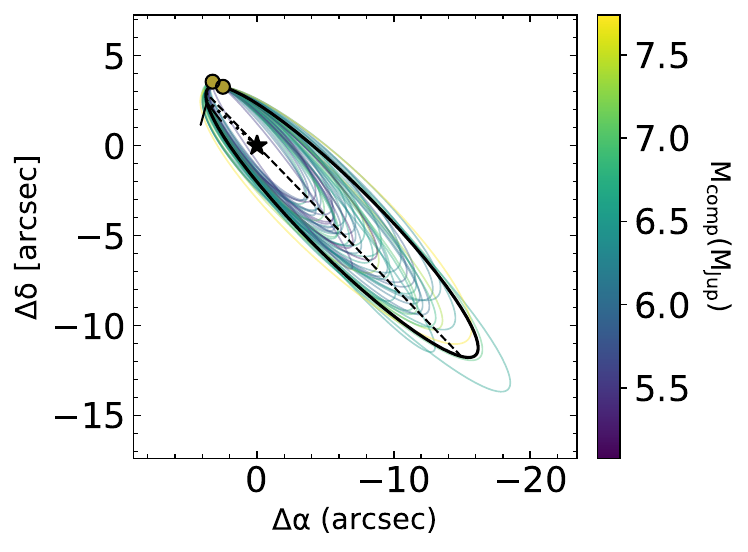}
    \includegraphics[width=0.37\textwidth]{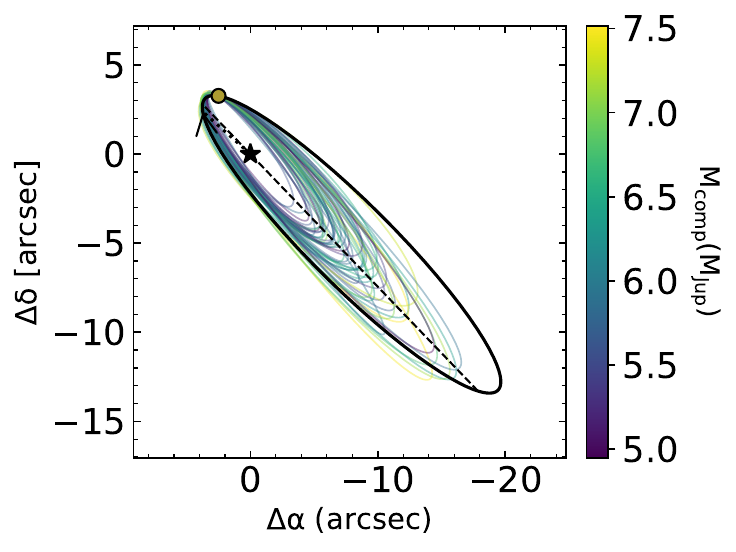}
    \includegraphics[width=0.37\textwidth]{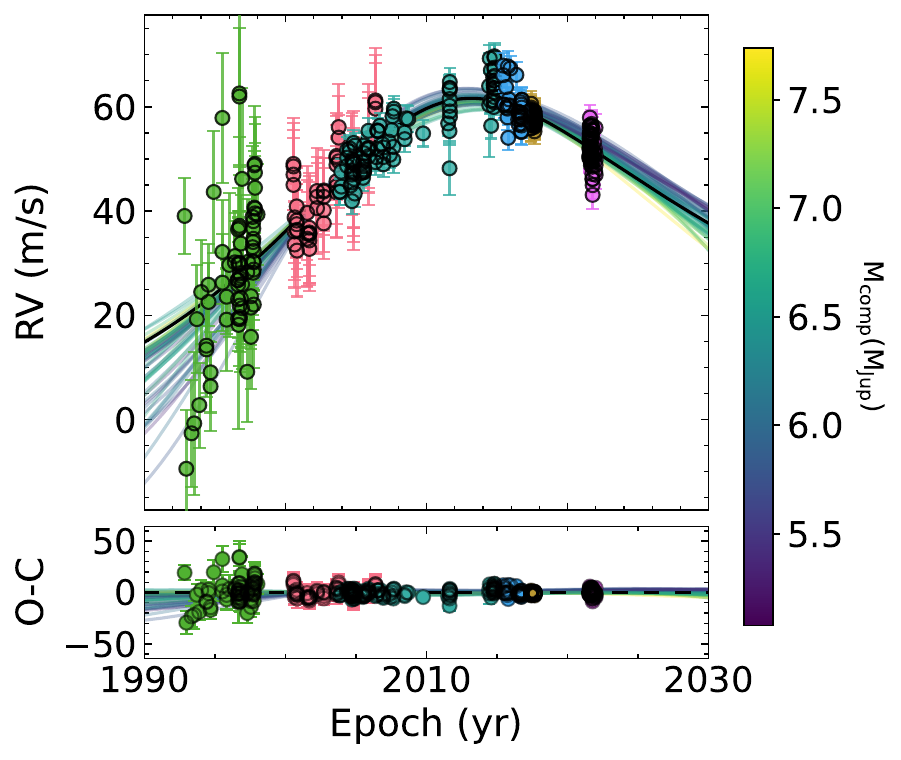}
    \includegraphics[width=0.37\textwidth]{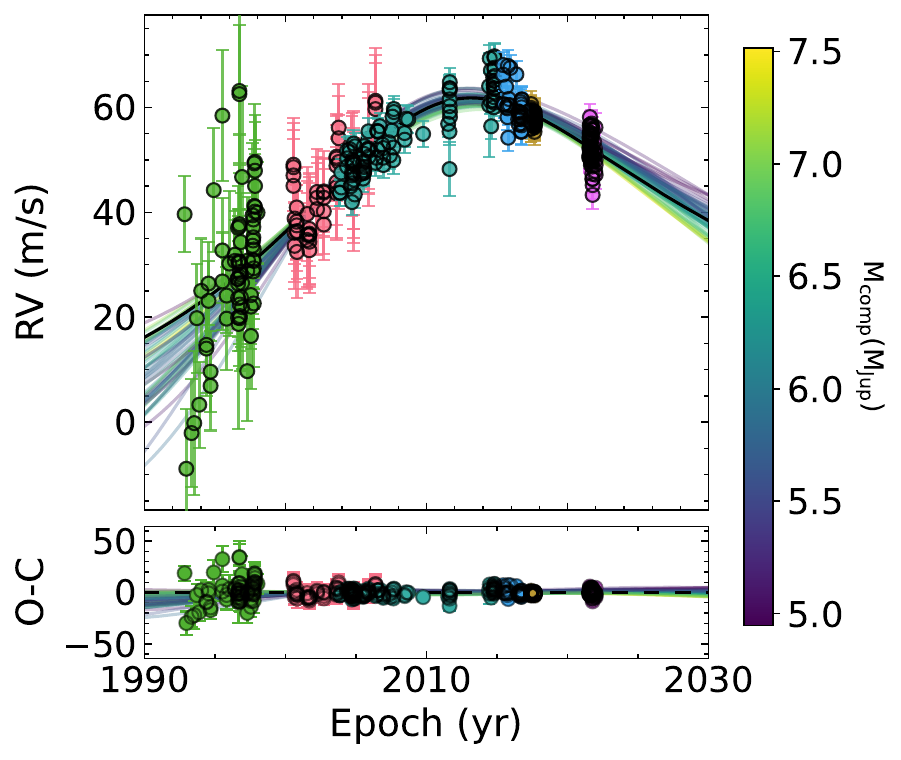}
    \includegraphics[width=0.37\textwidth]{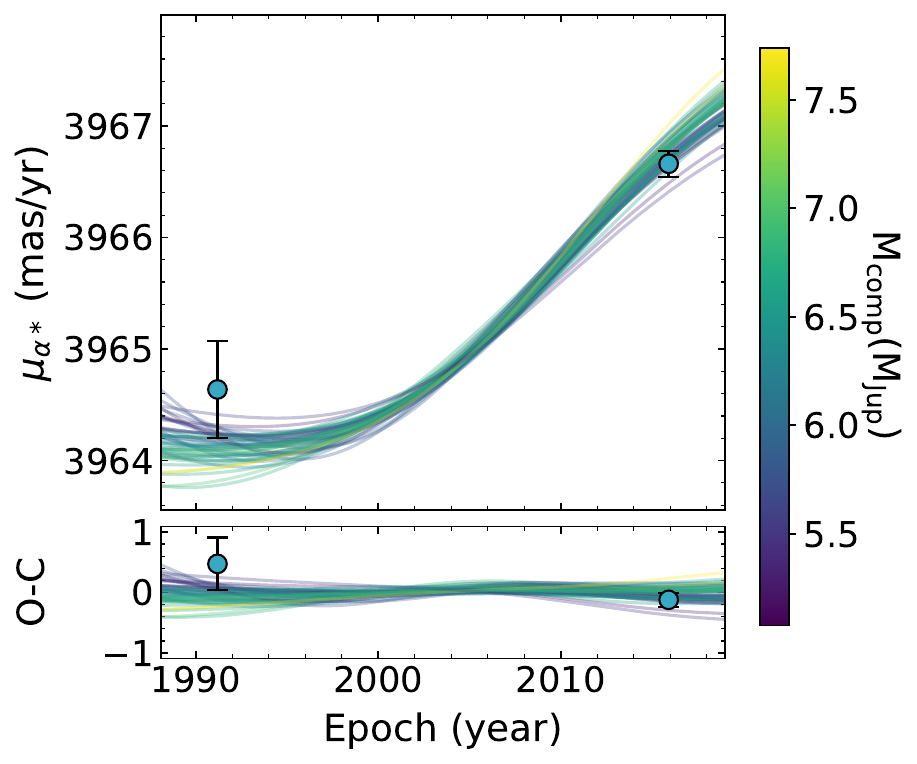}
    \includegraphics[width=0.37\textwidth]{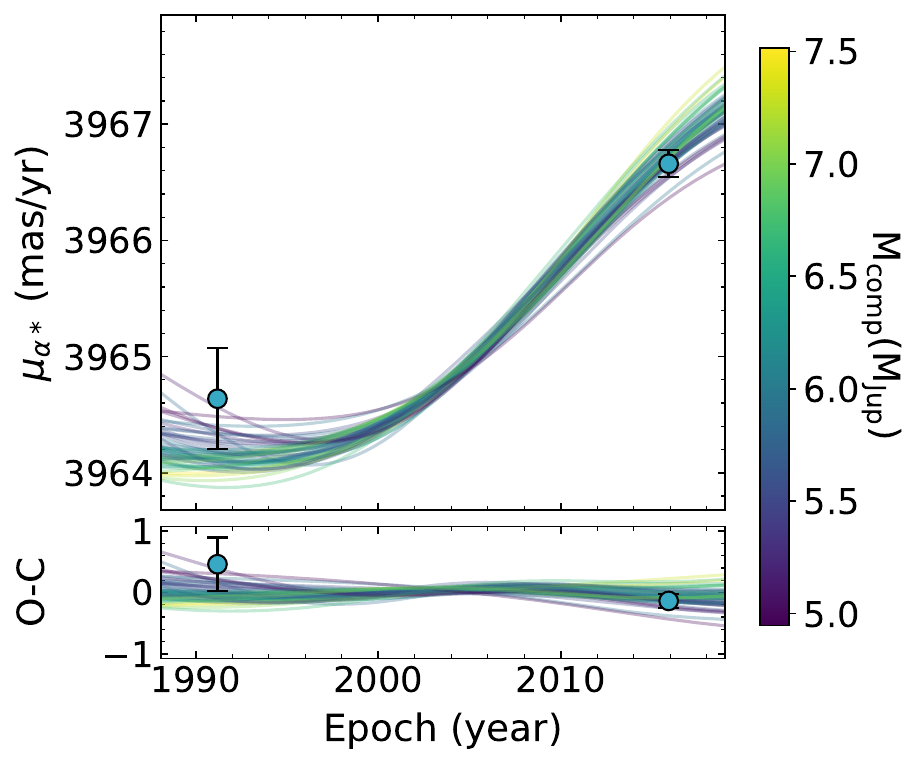}
    \includegraphics[width=0.37\textwidth]{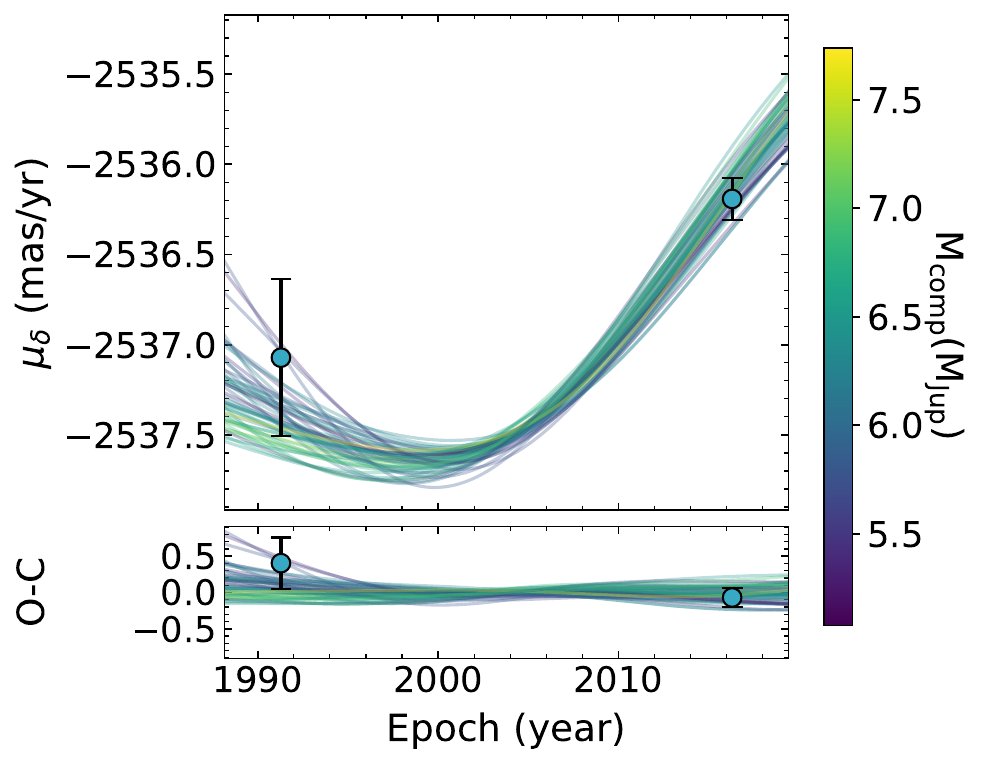}
    \includegraphics[width=0.37\textwidth]{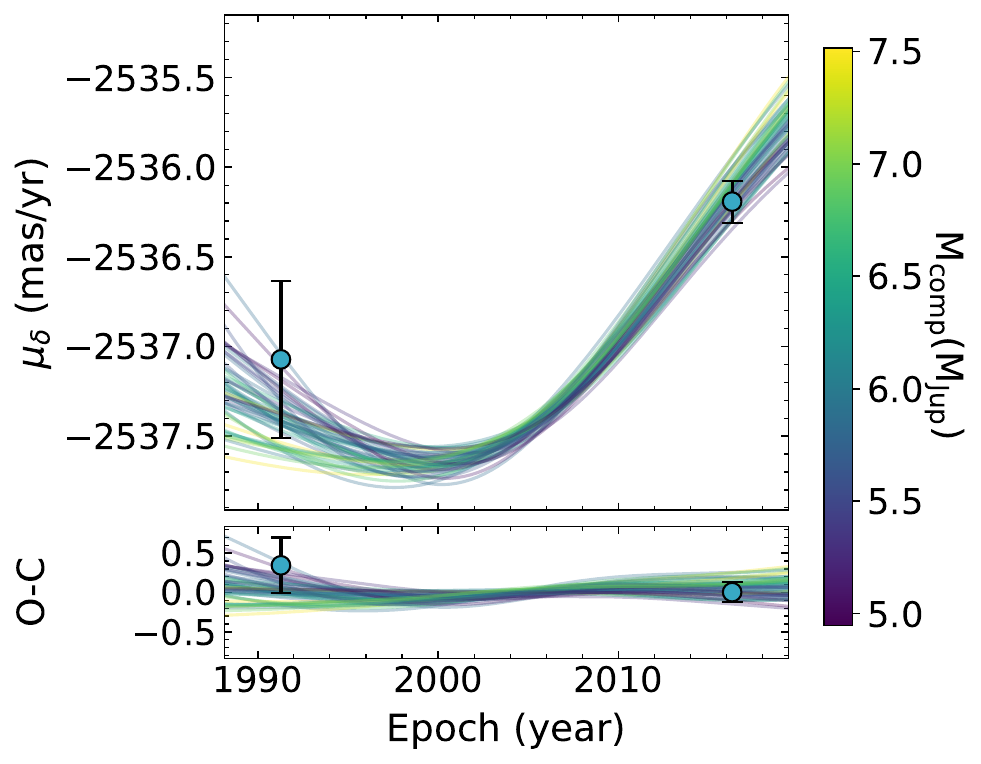}
    \caption{Orbit fits including RV, Hipparcos-Gaia astrometry, and both imaging points (left) or the JWST imaging only (right), assuming Eps Ind Ab is the only massive planet in the system. Rows represent (1) the on-sky orbit, (2) the RV of the target (3) the proper motion in right ascension of the host star and (4) the proper motion in declination of the host star. Orbits are very similar in both cases, and disagree strongly with the astrometric motion of the target. RVs are from LC (green), VLC (pink), HARPS03 (cyan), UVES (yellow), HARPS15 (blue) and HARPS20 (pink). Fits are well converged, and similar in both cases, but show a striking disagreement with the host star astrometry.}
    \label{fig:orbits}
\end{figure}

Our goal in targeting Eps~Ind~A with JWST coronagraphic imaging was to detect a known, massive planet for which dynamical mass information was accessible. However, the imaged planet is inconsistent with the previously claimed planet in the system: the planet has a significantly larger mass and semi-major axis than the claimed Eps~Ind~Ab\citep{Feng2019,Philipot2023,Feng2023}, and a strikingly different on-sky position than predicted. It is clear that the imaged planet imparts a significant, long-term RV acceleration on the host star, though it remains unclear whether this is the only planet in the system.

We attempted to fit the orbit of the imaged planet assuming that it is the only planet in the system, and is responsible for all the dynamical measurements (i.e.~RV and astrometry) of the host star. We used RV data from the CES long camera (LC) and very long camera (VLC)\citep{Zechmeister2013}, from UVES\citep{Feng2019} and from HARPS, using the data reduction presented in \citep{Trifonov2020}. Eps~Ind~A has been targeted intensively with HARPS, and we binned any datapoints collected within a single night with HARPS. We treated the HARPS data as three separate instruments, to account for any possible baseline shifts during an instrument upgrade in 2015 and a telescope shutdown in 2020. This totals 493 datapoints, from six independent instruments, covering the time period 1992-11-03 to 2021-12-29. We fitted orbits using \texttt{orvara}\citep{BrandtOrvara}, considering the 493 RV points, the two direct imaging epochs and the Hipparcos-Gaia astrometry of the host star, and using a host star mass prior of $0.76\pm0.04$M$_\odot$. This process converges to an orbit with mass ${6.31}_{-0.56}^{+0.60}$\mj, semi-major axis ${28.4}_{-7.2}^{+10}$au, eccentricity ${0.40}_{-0.18}^{+0.15}$, and inclination ${103.7}_{-2.3}^{+2.3}\degree$ (see Extended Data Figure \ref{fig:orbits}). Without the NEAR astrometry, we find an orbit with mass ${6.17}_{-0.57}^{+0.65}$\mj, 
semi-major axis ${31}_{-10}^{+13}$au, eccentricity ${0.47}_{-0.23}^{+0.15}$, and inclination ${104.0}_{-2.4}^{+2.4}\degree$. These orbits are highly consistent with eachother, further validating that the NEAR epoch is a positive redetection of the companion. These orbits are also consistent with all in-hand dynamical data. Curiously, several previous works had derived properties of the claimed planet Eps~Ind~Ab, and found consistent results\citep{Feng2019,Feng2023,Philipot2023}, but these results are inconsistent with the planet observed in this work. This may be due to over-fitting of the in-hand data since fitting accurate orbits with insufficient orbital phase coverage is notoriously hard\citep{Lagrange2023}, or may hint at an additional component in the system which biased the previous one-planet fits. The consistency between the fits with and without the NEAR astronomy points towards a one-planet system. We leave a detailed exploration of the discrepancy between the imaged planet and the previous RV solutions to a future work, but note that the dynamical mass derived in this work is only valid if Eps~Ind~Ab is the only massive planet present.

\subsection{Stellar Photometry} \label{methods:stellarphotometry}

\begin{figure}
\centering
    \includegraphics[width=0.5\textwidth]{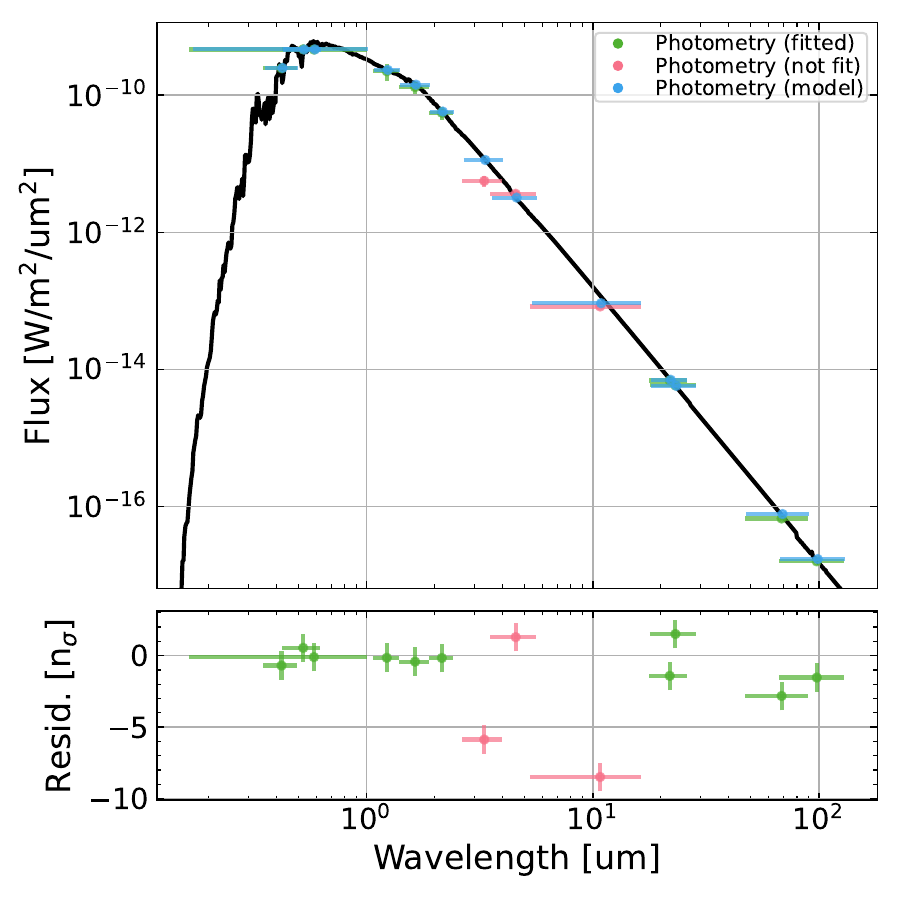}
    \caption{Best-fit spectrum for Eps~Ind~A. The spectrum is shown in black and integrated to R$\sim$100. Measured photometry is shown in green if included in the fit, and pink otherwise, while the model photometry is indicated in blue. The WISE W1, W2, and W3 channels are saturated, and in particular the W1 and W3 magnitudes\citep{Cutri2012} are several sigma fainter than the best-fit model.}
    \label{fig:starspec}
\end{figure}

We used BT-NextGen models \citep{Allard2011,Allard2012} to predict the photometry of the host star Eps~Ind~A in the JWST/MIRI and VLT/NACO filters. We fit spectra to the in-hand photometry from TYCHO, Gaia, 2MASS, WISE, Spitzer/MIPS and Herschel/PACS \citep{Hog2000,gaia_dr3,Skrutskie2006,Cutri2012,Gaspar2013}. We allowed effective temperature, $\log(g)$, metallicity, and radius to vary and held the stellar distance fixed at 3.639pc, and sampled the posterior with an MCMC chain (using \texttt{emcee}\citep{ForemanMackey2013}) to derive the best-fit spectrum and uncertainty. Best-fit values are T$_\textrm{eff}$=4760$\pm$15K, $\log(g)$=5.25$^{+0.18}_{-0.34}$; [Fe/H]=0.22$\pm$0.12 and R=0.679$\pm$0.004R$_\odot$, though we caution that these values are driven by photometry only, and spectroscopic fits of the stellar parameters should be used over these values when interpreting the metallicity and $\log(g)$ of the host star. We then derived model photometry and uncertainties by selecting samples from the MCMC posterior and integrating over the filter profiles. The best-fit spectrum, alongside the measured (green, pink) and model (blue) photometry of Eps~Ind~A is shown in Extended Data Figure \ref{fig:starspec}. We used this model to derive all companion photometry and flux limits in this work, including re-converting archival contrast curves to flux upper limits with this model.
The WISE observations of Eps~Ind~A appear to be unreliable. Eps~Ind~A is significantly saturated in the WISE W1, W2 and W3 observations (the WISE source catalog\citep{Cutri2012} lists saturation fractions of \texttt{w1sat}=0.253; \texttt{w2sat}=0.239, and \texttt{w3sat}=0.144). For this reason we do not include the W1, W2, and W3 magnitudes directly as part of our fitting process. Our derived stellar magnitudes show some discrepancy with those used to derive previous L' limits\citep{Viswanath2021}, which used the WISE W1 magnitude of 2.9 as a proxy for the L' magnitude of Eps~Ind~A; our stellar fit indicates a somewhat brighter magnitude for the star of 2.119$\pm$0.008 and 2.115$\pm$0.009 for the W1 and L' filters respectively. The only published L-band magnitude for Eps~Ind is L=2.12\citep{Morel1978}, consistent with our derived W1 and L' magnitudes.

\bibliography{eps-ind-bib}

\end{document}